\documentclass[twocolumn,aps,prx,superscriptaddress,floatfix]{revtex4-2}

\usepackage[strings]{underscore}
\usepackage{makecell}
\usepackage{array}

\newcommand{\mr}{\mathrm}
\newcommand{\mc}{\mathcal}

\usepackage{amssymb}
\usepackage{amsmath,bm}
\usepackage{graphicx}
\usepackage[normalem]{ulem}
\usepackage{bbding}
\usepackage{booktabs}
\usepackage[usenames, dvipsnames]{xcolor}
\usepackage{lipsum}
\usepackage{multirow}
\usepackage{url}
\usepackage{ulem}
\usepackage{tabularx}

\usepackage[colorlinks,
bookmarks=true,
linkcolor=blue,
urlcolor=blue,
anchorcolor=black,
citecolor=blue
]{hyperref}

 \renewcommand{\sout}[1]{}
\begin{document}

\title{Neutron Star versus Quark Star in the Multimessenger Era}

\author{Zheng Cao}
\affiliation{State Key Laboratory of Dark Matter Physics, Key Laboratory for Particle Astrophysics and Cosmology (MOE), and Shanghai Key Laboratory for
Particle Physics and Cosmology, School of Physics and Astronomy,
Shanghai Jiao Tong University, Shanghai 200240, China}
\affiliation{Tsung-Dao Lee Institute, Shanghai Jiao Tong University, Shanghai 201210, China}

\author{Lie-Wen Chen}
\thanks{Corresponding author}
\email{lwchen$@$sjtu.edu.cn}
\affiliation{State Key Laboratory of Dark Matter Physics, Key Laboratory for Particle Astrophysics and Cosmology (MOE), and Shanghai Key Laboratory for
Particle Physics and Cosmology, School of Physics and Astronomy,
Shanghai Jiao Tong University, Shanghai 200240, China}

\date{\today}

\begin{abstract}
Neutron stars (NSs), which may contain exotic degrees of freedom in their cores, and self-bound quark stars (QSs), composed entirely of absolutely stable deconfined quark matter, remain the two primary candidates for the compact objects  observed as pulsars and in gravitational-wave (GW) events from binary mergers.
Using a physics-agnostic equation of state (EOS) for compact star (CS) matter constructed via nonparametric Gaussian process regression, we perform Bayesian inference of the properties of NSs and QSs by analyzing multimessenger observations from GW170817, PSR J0740+6620, PSR J0030+0451, PSR J0437$-$4715, and PSR J0614$-$3329 together with {\it ab initio} calculations from perturbative quantum chromodynamics and chiral effective field theory.
We consider competing cases arising from three scenarios: the NS scenario, in which all CSs are NSs; the udQS scenario, in which all CSs are up-down QSs; and the two-family scenario, in which NSs and strange QSs (SQSs) coexist.
Our results based on systematic Bayesian model selection indicate that the most favored case belongs to the two-family scenario, with a Bayes factor of $6.1$ relative to the udQS case, corresponding to moderate evidence.
In this case, PSR J0740+6620 is favored as an SQS, while the other five CSs, including the two in GW170817, are favored as NSs, although the favored classification of PSR J0030+0451 further depends on the hot-spot model adopted in the NICER analysis.
The corresponding maximum mass of a static CS is $M_\text{TOV} = 1.72_{-0.16}^{+0.18} M_\odot$ for NSs and $M_\text{TOV} = 2.63_{-0.31}^{+0.36} M_\odot$ for SQSs at the $68\%$ confidence interval.
Moreover, the inferred sound speed in both NS and QS matter increases monotonically and saturates at high densities, whereas a pronounced peak structure is found in the NS scenario.
Our work provides a new framework for identifying the nature of individual CSs and constraining the properties of dense matter in both NSs and QSs.
In particular, the most favored case provides potential astrophysical support for the absolute stability of strange quark matter according to the Witten hypothesis.

\end{abstract}

\maketitle

\section{Introduction}
Understanding the nature of compact stars~(CSs) observed as pulsars and in gravitational-wave~(GW) events from binary mergers is one of the central problems in contemporary nuclear physics, particle physics, astrophysics, and cosmology.
The baryon number density in the cores of CSs can reach several times the nuclear saturation density ($ n_0 \equiv 0.16 \text{ fm}^{-3}$), making the CSs unique natural laboratories for exploring the properties of dense nuclear matter and the QCD phase diagram under extreme conditions of high density and low temperature~\cite{Lattimer:2004pg,Web05,Alford:2007xm,Fukushima:2010bq,Baym:2017whm,Blaschke:2018mqw,Burgio:2021vgk,Li:2021thg,Sedrakian:2022ata}, which are inaccessible in terrestrial experiments.
Theoretically, determining the properties of dense nuclear matter at high densities and low temperatures from {\it ab initio} calculations remains a major challenge due to the complicated nonperturbative nature of QCD~\cite{Brambilla:2014jmp}, and the internal composition (and thus the nature or type) of CSs remains largely uncertain.

Several competing scenarios, mainly including the neutron star (NS) scenario, the quark star (QS) scenario, and the two-family (TF) scenario, have been proposed to describe the astrophysical phenomena and the nature of CSs (see, e.g., Ref.~\cite{Web05}).
In the NS scenario, which is perhaps the most popular scenario for CSs, all CSs are assumed to be NSs, which include a liquid core and a solid crust, and their cores may contain not only neutrons and protons but also exotic degrees of freedom, such as hyperons, $\Delta$ resonances, meson condensates, and even deconfined quark matter. Note that in some literature, NSs containing exotic degrees of freedom may also be called hybrid stars.

According to the Witten hypothesis (also called the Bodmer-Witten or Bodmer-Witten-Terazawa hypothesis)~\cite{Bodmer:1971we,Witten:1984rs,Terazawa:1989iw,Farhi:1984qu}, the deconfined strange quark matter~(SQM), composed of $u$, $d$, and $s$ quarks with leptons, could be the true ground state of QCD matter. Under this hypothesis, CSs would be self-bound strange quark stars (SQSs) composed entirely of absolutely stable SQM.
Furthermore, building on the formation history of CSs, it is very likely that not all CSs are SQSs, because converting an NS formed in a core-collapse supernova into an SQS requires extreme conditions---principally the production of sufficient strangeness. Accordingly, the two-family scenario has been proposed~\cite{Bombaci:2000cv,Berezhiani:2002ks,Bombaci:2004mt,Drago:2004vu,Drago:2013fsa,Drago:2014oja,Drago:2015cea,Drago:2015dea,Bombaci:2016xuj,Wiktorowicz:2017swq,Bombaci:2020vgw,Drago:2020gqn,DiClemente:2024lzi}, in which two distinct branches of CSs coexist: an NS branch and an SQS branch.

More recently, it has been suggested that up-down quark matter~(udQM)---consisting solely of $u$ and $d$ quarks together with leptons---may replace SQM as the ground state of strong interaction matter~\cite{Holdom:2017gdc}. In this up-down quark star (udQS) scenario, all CSs are assumed to be up-down quark stars (udQSs), since only light quark flavors are involved.
We note that udQSs may also coexist with NSs~\cite{Zhang:2019mqb} if the transition rate is slow~\cite{Ren:2020tll}; however, such a mixed case would be practically equivalent to the TF scenario in the present work, so here we focus on the pure udQS case.
Within the above-discussed three scenarios, a CS could therefore be an NS (in the NS and TF scenarios), an SQS (in the TF scenario), or a udQS (in the udQS scenario).

The significant progress in both multimessenger observations of CS properties and {\it ab initio} theoretical inputs on the equation of state~(EOS) of dense matter may help to identify the type of individual CS and to test which scenario is most favored among the NS, udQS and TF scenarios for CSs.
On the observational side, substantial progress has been achieved in recent decades, driven by the rapid development of astrophysical facilities.
To date, simultaneous mass and radius measurements have been made via Neutron Star Interior Composition Explorer~(NICER) hot spot modeling for five pulsars: PSR J0740+6620~\cite{Miller:2021qha,Riley:2021pdl,Salmi:2024aum} with $M \sim 2M_{\odot}$, PSR J0030+0451~\cite{Riley:2019yda,Miller:2019cac,Vinciguerra:2023qxq}, PSR J0437$-$4715~\cite{Reardon:2024rdv,Choudhury:2024xbk} and PSR J0614$-$3329~\cite{Mauviard:2025dmd} with masses $M \sim 1.4M_{\odot}$ as well as PSR J1231$-$1411~\cite{Salmi:2024bss} with mass $M \sim 1.0M_{\odot}$.
In particular, the GW event GW170817~\cite{LIGOScientific:2017vwq,LIGOScientific:2018hze}, involving a binary compact-star merger and detected by the LIGO Scientific and Virgo Collaborations, marks the beginning of a new era of multimessenger astronomy and provides critical constraints on the tidal deformability of CSs.
On the theoretical side, {\it ab initio} calculations of dense matter have also advanced significantly in recent years.
At low densities, chiral effective field theory~(ChEFT), as a low-energy realization of QCD, provides reliable constraints on the EOS of NS matter up to densities $n \approx 1\sim 2n_0$, with controllable uncertainties~\cite{Hebeler:2013nza,Wellenhofer:2015qba,Keller:2022crb}.
At asymptotically high densities, where the baryon chemical potential reaches several GeV, perturbative QCD~(pQCD) calculations become feasible, and accordingly stringent constraints on the EOS of extremely dense matter (with baryon density of $ \sim 40 n_0 $) have been obtained (see, e.g., Ref.~\cite{Gorda:2021znl}).

Based on multimessenger observations and state-of-the-art {\it ab initio} calculations from ChEFT and pQCD, crucial constraints on the properties of dense matter and NSs have been obtained using physics-agnostic EOSs, under the assumption that all observed CSs are NSs, i.e., for the single case belonging to the NS scenario~\cite{Annala:2017llu,Annala:2019puf,Annala:2021gom,Gorda:2022jvk,Annala:2023cwx,Altiparmak:2022bke,Mroczek:2023zxo,Somasundaram:2022ztm,Han:2021kjx,Han:2022rug} (see also, e.g., Refs.~\cite{Capano:2019eae,Dietrich:2020efo,Pang:2021jta,Huth:2021bsp,Raaijmakers:2021uju,Tiwari:2023tkj,Brandes:2022nxa,Tsang:2023vhh} for complementary studies that do not incorporate pQCD constraints).
Moreover, the case in which all observed CSs are udQSs has been investigated within specific quark matter models~\cite{Zhao:2019xqy,Zhang:2019mqb,Cao:2020zxi,Ren:2020tll,Wang:2025lwv}.
Some cases belonging to the TF scenario have also been studied, with the NS or SQS identity of each observed CS fixed  {\it a priori} through phenomenological arguments~\cite{Traversi:2020dho,Traversi:2021fad}.
In all of these studies, the NS or QS nature of each CS is assumed {\it a priori} rather than determined from the data, and a systematic comparison of all competing cases is still lacking.
A comparative Bayesian analysis of all competing cases, with minimal model assumptions about the EOS of dense matter, is therefore particularly informative, as it allows one to infer the nature of each individual CS and the properties of dense matter in a model-independent manner.
Here we report results of the first work in this direction.

In this paper, we perform a Bayesian inference of the EOS properties together with a systematic Bayesian model selection (according to the Bayesian evidence~\cite{Jeffreys:1939xee, Trotta:2008qt,Lee14}) among all competing cases spanning the NS, udQS, and TF scenarios.
Using a physics-agnostic EOS for CS matter constructed via nonparametric Gaussian process (GP) regression and combining the current multimessenger data with state-of-the-art {\it ab initio} constraints from ChEFT and pQCD,
we find that the most favored case belongs to the TF scenario, with a Bayes factor of $6.1$ indicating moderate evidence.
In this case, the two CSs in GW170817, as well as PSR J0030+0451, PSR J0437$-$4715, and PSR J0614$-$3329, are favored as NSs, whereas PSR J0740+6620 is favored as an SQS.
Because the NICER mass-radius constraint for PSR J0030+0451 depends on the adopted hot-spot model, we also examine alternative treatments of this source and find that its favored classification may change with the treatment although the classification of other CSs remains unchanged.
Moreover, our results suggest case-dependent density behaviors of the sound velocity and trace anomaly in CS matter.

The most favored case belonging to the TF scenario may also carry possible implications for the nature of SQM and even dark matter.
Because the coexistence of NSs and self-bound SQSs requires SQM to be absolutely stable, the moderate evidence for this case provides potential support for the Witten hypothesis. Within this picture, cosmological strangelets have been proposed as possible baryonic dark-matter candidates~\cite{DiClemente:2024lzi}.

\section{Methods}
\subsection{Physics-agnostic EOS}
In this work, we construct the physics-agnostic EOSs for NS and QS matter using the nonparametric GP regression method.
For NS matter, we adopt an approach similar to that recently introduced in Ref.~\cite{Gorda:2022jvk}. By expanding interactions to next-to-next-to-next-to-leading order, ChEFT provides reliable calculations for the EOS of NS matter up to about 0.25 fm$ ^{-3} $  ~\cite{Keller:2022crb}.
Consequently, for the NS matter EOS within the density interval $[0.58 n_0, 1.5 n_0]$, we condition the GP on the ChEFT results and then extrapolate it to $ 12 n_0$ using the conditioned GP.
For densities below the lower bound ($n_1 = 0.58n_0$) of the GP domain, the BPS EOS~\cite{Baym:1971pw} is employed.
We have checked that using other crust EOSs, such as SFHo~\cite{Steiner:2012rk} and HS2020~\cite{Shen:2020sec}, would not change our conclusion.

In the GP regression, an auxiliary variable $\phi(n)$ is defined as $\phi(n)=-\ln \left(1 / c_s^2(n)-1\right)$, which is modeled as a Gaussian process:
\begin{equation}
    \phi(n) \sim \mathcal{N}\left(-\ln \left(1 / \bar{c}_s^2-1\right), K\left(n, n^{\prime}\right)\right)
\end{equation}
where {$n$ is the baryon number density, $c_s^2(n)$ is the squared sound speed,} $K\left(n, n^{\prime}\right)$ represents the kernel function, chosen as $K\left(n, n^{\prime}\right) = \eta e^{-\left(n-n^{\prime}\right)^2 / 2 l^2}$.
The prior distributions for the hyperparameters  $l$, $\eta$, and $\bar{c}_s^2$ are taken as those used in Ref.~\cite{Gorda:2022jvk}:
\begin{equation}
\begin{aligned}
l & \sim \mathcal{N}\left(1.0 n_0,\left(0.25 n_0\right)^2\right), \quad \eta \sim \mathcal{N}\left(1.25,0.2^2\right), \\
\bar{c}_s^2 & \sim \mathcal{N}\left(0.5,0.25^2\right) .
\end{aligned}
\end{equation}
We have verified that our conclusions are unaffected by adopting wider hyperpriors, $l \sim U(0.1 n_0, 4 n_0)$, $\eta \sim U(0.1, 10)$, and $\bar{c}_s^2 \sim U(0, 1)$.
With the hyperparameters specified, the GP for the NS matter EOS is conditioned on ChEFT within the density interval [0.58, 1.5]$ n_0 $.
Specifically,  the average of the squared sound speeds $c_s^2$ from the stiffest and softest ChEFT EOSs is taken as the GP mean, and their difference as the 90\% confidence interval (CI). Subsequently, $\phi(n)$ is sampled from the conditioned GP, and the full EOS of NS matter up to $ 12 n_0 $  is obtained using the fundamental thermodynamic relations:
\begin{align}
	\mu(n)&=\mu_1 \exp \left[\int_{n_1}^n d n^{\prime} \frac{c_s^2\left(n^{\prime}\right)} {n^{\prime}}\right], \label{Theo1}\\
	\varepsilon(n)&=\varepsilon_1+\int_{n_1}^n d n^{\prime} \mu\left(n^{\prime}\right),  \label{Theo2}\\
	p(n)&=-\varepsilon(n)+\mu(n) n, \label{Theo3}
\end{align}
where $\mu(n)$, $\varepsilon(n)$ and $p(n)$ are, respectively, the chemical potential, energy density, and pressure of the CS matter at baryon number density $n$, $n_1$ is the density lower bound of the GP domain~(i.e., $0.58n_0$ for NS matter EOS construction), and $\mu_1$~($\varepsilon_1$) is the corresponding chemical potential~(energy density) at $n_1$.

As for the QS matter EOS, we construct it directly from its self-bound nature, which requires the pressure to vanish at a finite surface density $n_1$.
Accordingly, the baryon number density and chemical potential at this zero-pressure surface are treated as free parameters of the EOS.
We sample the chemical potential $\mu_1$ at the zero-pressure-point density $n_1$ uniformly within the interval $[400, 930]$~MeV.
The upper limit $930$~MeV ensures that $\mu_1$ remains below the energy per baryon of observed stable nuclei, thereby satisfying the condition of absolute stability for the quark matter~\cite{Bodmer:1971we,Witten:1984rs,Terazawa:1989iw,Farhi:1984qu}.
The surface density $n_1$ is sampled from a uniform distribution over the interval $[0.5, 4]n_0$.
The energy density at $n_1$ is then determined according to the thermodynamic relation $\varepsilon_1 = \mu_1 n_1$.
As we will see later, the prior intervals $\mu_1 = [400, 930]$~MeV and $n_1 = [0.5, 4]n_0$ are large enough to ensure the convergence of our results.
With $\phi(n)$ sampled directly from the GP prior distribution, the full EOS for QS matter up to $12n_0$ is then obtained using the thermodynamic relations Eqs.~\eqref{Theo1}, \eqref{Theo2} and \eqref{Theo3}.

\subsection{Bayesian analysis}
We use a Bayesian hierarchical model to combine constraints from multiple observations with uncertainties and estimate the EOS parameters for NS and QS matter.
According to Bayes' theorem, for the given dataset $  \vec{d}$ and hypothesis $  \mathcal{H} $, the posterior distribution of EOS parameters $  \theta $ can be written as (see, e.g., Refs.~\cite{Thrane:2018qnx})
\begin{equation}\label{Bayes}
    p(\theta | \vec{d},\mathcal{H}) = \frac{\prod_i  \mathcal{L}(d_i|\theta,\mc{H})\pi (\theta | \mc{H})}{\mc{Z}_{\mc{H}}(\vec{d})},
\end{equation}
where $i$ runs over individual constraints and each constraint is independent of the others, $  \pi(\theta| \mc{H} )$ is the hyper-prior distribution for $  \theta $ and here is chosen as a uniform distribution,
$\mathcal{L}(d_i| \theta,\mc{H}) $ is the likelihood of the EOS parameters under the assumption of $ \mc{H} $  for data $  d_i $.
The denominator of Eq.~\eqref{Bayes}
\begin{equation}
\mc{Z}_{\mc{H}}(\vec{d}) \equiv \int \prod_i \mathcal{L}(d_i | \theta,\mc{H})\pi (\theta | \mc{H})  \mr{d} \theta
\end{equation}
is a normalization factor called the \textit{evidence}, which quantifies how well the hypothesis is supported by the data.
Based on the dataset $ \vec{d} $,
the Bayes factor  $ \mathcal{B}^\text{1}_\text{2}  $  for hypothesis $1$ against {hypothesis} $2$ can be obtained as~\cite{Jeffreys:1939xee}
\begin{equation}\label{BF}
	 \mathcal{B}^\text{1}_\text{2}   = \mc{Z}_\text{1}(\vec{d})/ \mc{Z}_\text{2}(\vec{d}).
\end{equation}
The Bayes factor has been extensively adopted in many EOS-related studies, e.g., selecting EOSs~\cite{Yuan:2024hge,Kashyap:2025cpd} and studying the phase transition inside CSs~\cite{Pang:2021jta,Brandes:2023hma,Cao:2026dwk}.
A value of $\mathcal{B}^\text{1}_\text{2} = 1$ implies that the observational data $ \vec{d} $ support both hypotheses equally.
Furthermore, values of $\mathcal{B}^\text{1}_\text{2} $ within the intervals $[1,3]$, $[3, 10]$, $[10, 30]$, $[30, 100]$, and $[100, +\infty)$ correspond to anecdotal, moderate, strong, very strong, and extreme evidence, respectively, in favor of hypothesis 1~\cite{Jeffreys:1939xee, Trotta:2008qt,Lee14}.

For the NICER data $ d_\text{MR}$ on the simultaneous mass-radius measurement,
the likelihood can be written as
\begin{equation}
	\mathcal{L}(d_{\rm MR} | \theta)=\int \mathrm{d}m \mathcal{L}[d_{\rm MR} | m,R(m,\theta)] \pi(m|\theta),
\end{equation}
{where $m$ and $R$ represent the mass and radius of CSs, respectively,} and $\pi(m|\theta)$ is a uniform prior on the mass.
	We use Kernel Density Estimation~(KDE) to approximate the experimental mass-radius distribution~\cite{Vinciguerra0451,Salmi6620,Choudhury4715, Lucien0614}.

Turning to gravitational-wave events, the corresponding likelihood takes the form (see, e.g.,~\cite{HernandezVivanco:2019vvk})
\begin{equation}
\begin{aligned}
	\mathcal{L} (d_{\text{GW}} | \theta) &=  \int \mathrm{d}m_1\, \mathrm{d}m_2\,  \pi(m_1, m_2 | \theta) \\
	& \times \mathcal{L}[d_{\text{GW}} | m_1, m_2, \Lambda_1(m_1, \theta), \Lambda_2(m_2, \theta)],
\end{aligned}
\end{equation}
where $\mathcal{L}[d_\text{GW} | m_1, m_2, \Lambda_1, \Lambda_2]$ is the nuisance-marginalized likelihood~\cite{HernandezVivanco:2020cyp} which has marginalized over extrinsic parameters of the source.
Here $m_1$ and $m_2$ ($m_1 \geq m_2$) denote the masses of the primary and secondary components of the binary, respectively, while the tidal deformabilities $\Lambda_1$ and $\Lambda_2$ are determined by the respective masses together with the EOS parameters $\theta$.
For GW170817, a low-spin prior is assumed, and the details of the prior distributions for the waveform can be found in Tables 1 and 2 of Ref.~\cite{HernandezVivanco:2020cyp}.

For the pQCD constraints, based on thermodynamic stability and causality, the results from pQCD at extremely high densities can be used to constrain the EOS in the intermediate density region corresponding to the NS interior by fully taking advantage of thermodynamic potentials~\cite{Komoltsev:2021jzg}.
At a high chemical potential $ \mu_H = 2.6 $ GeV (i.e. $n\approx 40n_0 $), the uncertainties of thermodynamic quantities could be parameterized by a dimensionless parameter $ X $ and could be expressed by a set $\vec{\beta}_{\mathrm{pQCD}}(X)=\left\{p_{\mathrm{pQCD}}\left(\mu_H, X\right), n_{\mathrm{pQCD}}\left(\mu_H, X\right), \mu_H\right\}$. The probability distribution for the quantities at $\mu_H$, $\vec{\beta}_{H} = \{p_H, n_H, \mu_H\}$, can be obtained by~\cite{Gorda:2022jvk}
\begin{equation}
	P(\vec{\beta}_H)=\int d(\ln X) w(\ln X) \delta^{(3)}\left(\vec{\beta}_H - \vec{\beta}_{\mathrm{pQCD}}(X)\right),
\end{equation}
with the weight function $ w $.
By integrating the uncertainties, one can obtain the corresponding likelihood~\cite{Gorda:2022jvk}
\begin{equation}
	{\mathcal{L}(\mathrm{pQCD} | \theta)=\int d \vec{\beta}_H P(\vec{\beta}_H) \mathbf{1}_{\left[\Delta p_{\min }, \Delta p_{\max }\right]}(\Delta p)},
\end{equation}
with $ \Delta p =p_\text{pQCD} - p_L $ , where $ p_L $ is the pressure of the last density point in the GP sampling for the dense matter EOS (i.e., the last density point is $12n_0 $ here).
The indicator function $ \mathbf{1}_S $ takes the value one for elements in the subset $ S $ and zero otherwise.

In the present Bayesian analyses, we consider six CSs precisely measured by NICER and the LIGO/Virgo Collaborations, namely: the most massive PSR J0740+6620~\cite{Salmi:2024aum}, the most recent mass-radius measurement of PSR J0030+0451 analyzed using the hot spot model \texttt{ST+PDT}~\cite{Vinciguerra:2023qxq}, PSR J0437$-$4715~\cite{Choudhury:2024xbk} and PSR J0614$-$3329~\cite{Mauviard:2025dmd}; the tidal deformability information of the primary component GW17\_1 ($M \in [1.36, 1.60]M_\odot$) and the secondary component GW17\_2 ($M \in [1.16, 1.36]M_\odot$) from GW170817~\cite{LIGOScientific:2017vwq,LIGOScientific:2018hze}; along with the pQCD constraints at ultra-high densities~\cite{Gorda:2021znl,Komoltsev:2021jzg,Gorda:2022jvk,LHQCD} (and ChEFT results for NSs~\cite{Keller:2022crb}).
These observational and theoretical constraints constitute the default dataset in the present work, denoted as $\vec{d}_\text{def}$.

It should be noted that for the mass-radius data of PSR J0030+0451, several hot spot models, including \texttt{ST+PST}, \texttt{ST+PDT}, and \texttt{PDT-U}, can describe the data well; we will investigate in Sec.~\ref{ImpactJ0030} how our results change when different datasets employing different hot spot models are used.
We also note that NICER has observed PSR J1231$-$1411; however, the hot spot geometry of this pulsar is complex.
To fit the observational data, Ref.~\cite{Salmi:2024bss} adopted the complex \texttt{PDT-U} case.
However, this model does not converge without an informed radius prior, and the final results vary from $R = 12.6 \pm 0.3$~km, when previous observations and EOS information are considered, to $R = 13.5_{-0.5}^{+0.3}$~km when a tight radius prior is assumed.
A separate analysis using the \texttt{STS-PST} hot spot model with a broad radius range yielded $R = 9.91_{-0.86}^{+0.88}$~km~\cite{Qi:2025mpn}.
Given the uncertainties of these results, we do not consider this observation in our present analysis.

Each CS scenario prescribes a specific classification of the six observed CSs, thereby defining a distinct set of competing cases for Bayesian comparison.
In the NS scenario, all six CSs are classified as NSs, whereas in the udQS scenario, they are all classified as udQSs.
Each single-family scenario thus yields a single case — referred to as the NS case and the udQS case, respectively.
Since all six CSs belong to the same stellar family, only one type of EOS — for NS matter or QS matter, respectively — is required to describe the properties of dense matter and the stellar structure.
Consequently, we sample one type of EOS to characterize all six CSs simultaneously in each single-family scenario.

In the TF scenario, however, the situation is more complicated.
Theoretically, this scenario contains two independent mass-radius branches: a branch of NSs and a branch of SQSs.
To model this, we  account for two distinct types of dense matter (i.e., NS matter and QS matter) simultaneously.
We combine the parameter spaces of the QS and NS matter EOSs, sampling pairs of EOSs, with one for NSs and the other for SQSs.
From an observational perspective, this implies that a specific CS could belong to either the NS branch or the QS branch, provided its mass-radius or mass-tidal deformability constraints are consistent with the respective theoretical predictions.
For the six CSs considered in the present work, this leads to a total of $2^6 = 64$ possible  cases.
For convenience, we employ the labels N (NS) and Q (QS) to denote the type~(NS or QS) of the sources in the set \{PSR J0740+6620, PSR J0030+0451, PSR J0437$-$4715, PSR J0614$-$3329, GW17\_1, GW17\_2\}.
For instance, the notation TF-NQNNNN indicates a case where PSR J0030+0451 is a QS, while the other five CSs are NSs.

Taken together, the three scenarios yield 66 competing cases: one for the NS scenario, one for the udQS scenario, and $2^6 = 64$ for the TF scenario.
Since only one of these cases can be realized in nature for the six CSs, we treat them as mutually exclusive hypotheses with an equal prior probability of $1/66$ for each, and Bayesian model selection is required to single out the one most favored by the data.
To this end, the Bayesian evidence and the associated Bayes factor of each case quantifies its relative support from current multimessenger observations and {\it ab initio} calculations, thereby providing a data-driven identification of the underlying CS classification.

Besides ranking the individual cases, the Bayesian evidences of all cases also allow us to compare the three scenarios as a whole through Bayesian model averaging. For this purpose, we simply assign an equal prior weight of $1/64$ to each case in the TF scenario, so that the model-averaged Bayes factor of the TF scenario relative to the udQS case is $\bar{\mathcal{B}}^\text{TF}_\text{udQS} = \sum_{a} \mathcal{B}^a_\text{udQS} / 64$, where $\mathcal{B}^a_\text{udQS}$ is the Bayes factor of case $a$ relative to the udQS case.

To evaluate the posterior distribution and Bayesian evidence, we sample $2 \times 10^6$ EOSs from the GP per case and assign each sample the corresponding likelihood weight $w_i$.
For an ensemble of $N$ EOSs, the evidence is estimated by averaging the weights, i.e., $\mc{Z}_{\mc{H}}(\vec{d}) = \left(\sum_i w_i\right)/N$.
To quantify the robustness of this Monte Carlo estimate, we compute the effective sample size
$ N_\text{eff} = \left(\sum_i w_i\right)^2/\sum_i w_i^2 $~\cite{liu2001monte}
for each case. The smallest effective sample size  among all cases is approximately 3000, sufficient to ensure the convergence and robustness of our results.

\section{Results and Discussions}
\label{sec:results}
\subsection{Results in the default dataset $\vec{d}_\text{def}$}
\subsubsection{Properties of compact stars}
Before presenting the posterior distributions, we first identify which of the 66 competing cases is most favored by current observational data and {\it ab initio} calculations.
We take the case in the udQS scenario  (i.e., the udQS case)  as the reference, whose Bayesian evidence is found to be $\mathcal{Z}_\text{udQS} = 0.032$, and compute the Bayes factor of every other case relative to the udQS case.
In Tab.~\ref{Tab-def-TFBF}, we list the first fifteen cases within the TF scenario that have the largest Bayes factors, together with the case in the NS case (i.e., the NS scenario).
From this table, one sees that the most favored case is TF-QNNNNN among the 66 competing cases, with the largest Bayes factor $\mathcal{B}^\text{TF-QNNNNN}_\text{udQS} = 6.1$, which constitutes moderate Bayesian evidence in support of the TF-QNNNNN case.
In this case, the heaviest CS, PSR J0740+6620, is favored as an SQS while the other five CSs are favored as NSs.
It should be noted that the Bayes factors of the top-ranked cases in Tab.~\ref{Tab-def-TFBF} differ by less than a factor of two and are thus not statistically distinguishable, but all of them share the assignment of PSR J0740+6620 as an SQS. In addition, taking all cases in the TF scenario into account with an equal prior weight of $1/64$, the model-averaged Bayes factor of the TF scenario as a whole is $\bar{\mathcal{B}}^\text{TF}_\text{udQS} = 1.2$.
By contrast, for the NS case we find $\mathcal{B}^\text{NS}_\text{udQS} = 1.0$, indicating that the NS and udQS cases are equally supported by  current astrophysical observations and {\it ab initio} calculations.
This result also implies that the current multimessenger data together with {\it ab initio} calculations cannot distinguish the single-family NS and udQS cases.

\begin{table}[htb!]
	\footnotesize
    \centering
    \caption{Bayes factors ($ \mathcal{B} $) relative to the case in udQS scenario for the fifteen top-ranked cases in the TF scenario. In the bottom row, the result for the case in the NS scenario is included for comparison. The CS types~(N or Q) of the six CSs in each TF case are sorted according to the sequence: \{PSR J0740+6620, PSR J0030+0451, PSR J0437$-$4715, PSR J0614$-$3329, GW17\_1, GW17\_2\}.}
    \label{Tab-def-TFBF}
	\begin{tabular}{|l|c|l|c|l|c|}
        \hline
        Case & Value & Case & Value & Case & Value \\
        \hline
        {\scriptsize TF-QNNNNN} & 6.1 & {\scriptsize TF-QNNQNN} & 4.4 & {\scriptsize TF-QQNNNN} & 4.1 \\
        \hline
        {\scriptsize TF-QNQNNN} & 3.6 & {\scriptsize TF-QQQNNN} & 3.4 & {\scriptsize TF-QNNQNQ} & 2.9 \\
        \hline
        {\scriptsize TF-QNQQNN} & 2.7 & {\scriptsize TF-QNNNNQ} & 2.6 & {\scriptsize TF-QNNQQN} & 2.5 \\
        \hline
        {\scriptsize TF-QNNNQN} & 2.5 & {\scriptsize TF-QQNQNN} & 2.1 & {\scriptsize TF-QQQQNN} & 2.1 \\
        \hline
        {\scriptsize TF-QNNQQQ} & 2.1 & {\scriptsize TF-NNNQQQ} & 1.9 & {\scriptsize TF-QNNNQQ} & 1.6 \\
        \hline
        \multicolumn{6}{|c|}{\scriptsize  $\mathcal{B}^\text{NS}_\text{udQS} = 1.0$} \\
        \hline
	\end{tabular}
\end{table}

\begin{figure*}[!htbp]
	\centering
	\includegraphics[width=0.95\linewidth]{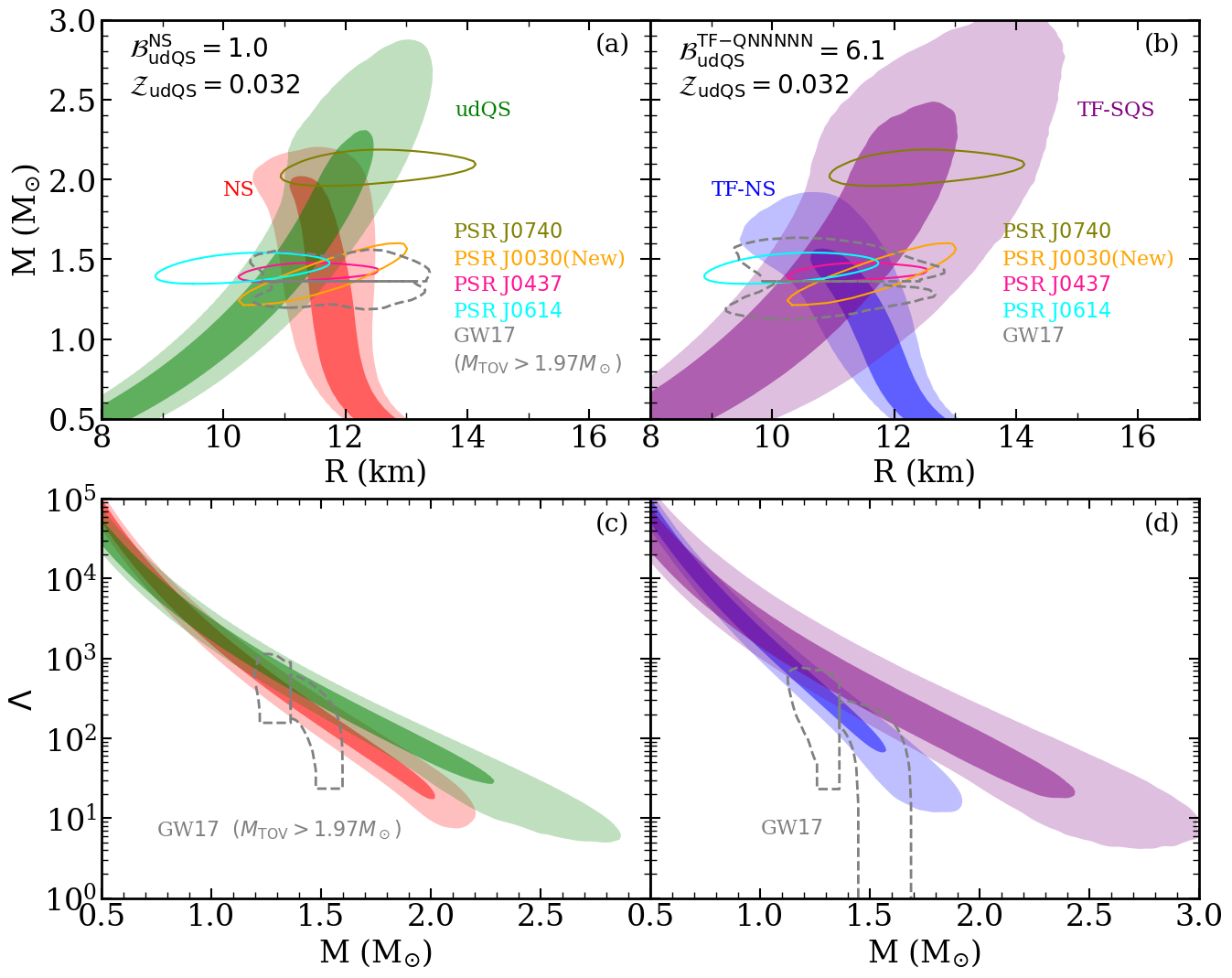}
	\caption{
    Posterior of mass-radius relation~(panels (a) and (b)) and tidal deformability~(panels (c) and (d)) in different  cases in the $\vec{d}_\text{def}$. The inner (outer) shaded region represents the 68\% (95\%) CI.
    For comparison, we also show the NICER mass-radius posterior distributions of PSR J0740+6620~\cite{Salmi:2024aum}, PSR J0030+0451~\cite{Vinciguerra:2023qxq}, PSR J0437$-$4715~\cite{Choudhury:2024xbk} and PSR J0614$-$3329~\cite{Mauviard:2025dmd}.
    The mass-radius and tidal deformability-mass relations for the binary components of GW170817 are also presented.
    These constraints are derived using parametrized EOSs subject to a maximum mass lower limit of 1.97 $M_\odot$ (panel (c)) and via EOS-insensitive relations without imposing a maximum-mass constraint~(panel (d))~\cite{LIGOScientific:2018cki}.
    Note that for GW170817, our Bayesian analysis directly uses the tidal information inferred without specific EOS assumptions.
    All of the experimental constraints are shown in 68\% CI.
    }
	\label{Fig-def-MR}
\end{figure*}

To understand why the NS and udQS cases are equally supported by the current observations and {\it ab initio} calculations, we present in Fig.~\ref{Fig-def-MR}(a) the posterior distributions of the mass-radius relations derived from $\vec{d}_\text{def}$ for both cases.
The corresponding data from NICER for the four CSs in the $\vec{d}_\text{def}$ are also included for comparison.
For GW170817, although our Bayesian analysis directly incorporates the tidal information (instead of the mass-radius constraints obtained therein), we also visualize the mass-radius relations for its two components. These regions are mapped from the tidal data via a  parametrized NS matter EOS  that imposes a lower limit of $1.97M_\odot$ on the NS maximum mass~\cite{LIGOScientific:2018cki}.
One sees from Fig.~\ref{Fig-def-MR}(a) that the udQS radius is restricted to relatively small values, overlapping only marginally with the lower~(left) radius bound of PSR J0030+0451.
Conversely, while the NS case aligns well with the central mass-radius constraint of PSR J0030+0451, it overlaps only with the upper~(right) radius bound of PSR J0614$-$3329.
In addition, around mass $M \sim 2M_\odot$, the udQS case describes the mass-radius constraint of PSR J0740+6620 better than the NS case, while the NS case appears to better describe GW170817. Overall,  both cases can describe the multimessenger observations only partially, with no obvious preference for either of the two cases, and hence the NS and udQS hypotheses are equally supported.

It is interesting to see that despite their equivalent Bayesian support, the NS and udQS cases predict markedly different macroscopic CS structure.
The udQSs have a higher maximum mass but a smaller canonical radius than NSs.
In particular, we note
$M_{\rm{TOV,udQS}} = 2.50_{-0.27}^{+0.30}M_\odot$ compared to $M_{\rm{TOV,NS}} = 2.09_{-0.08}^{+0.10}M_\odot$,
and $R_{\rm{1.4,udQS}} = 10.99_{-0.29}^{+0.33}~\text{km}$ compared to $R_{\rm{1.4,NS}} = 11.73_{-0.38}^{+0.40}~\text{km}$ (68\% CI).

Having established the single-family results, we now examine why the TF-QNNNNN case provides a superior global fit.
Fig.~\ref{Fig-def-MR}(b) shows the posterior mass-radius relations of the NS branch (TF-NS) and the SQS branch (TF-SQS) for the most favored  case, TF-QNNNNN.
For comparison, the  multimessenger constraints are also displayed.
For GW170817, however, we plot in Fig.~\ref{Fig-def-MR}(b) the mass-radius constraints derived from EOS-insensitive relations without imposing the lower limit of $1.97M_\odot$ on the maximum NS mass~\cite{LIGOScientific:2018cki}, since PSR J0740+6620 is not favored as an NS in the TF-QNNNNN case and the NS branch is therefore not required to satisfy $M \geq 1.97M_\odot$.

As shown in Fig.~\ref{Fig-def-MR}(b), the TF-QNNNNN case yields global stellar properties distinct from those in the NS and udQS cases, with a softer NS branch and a stiffer SQS branch that supports a higher maximum mass.
For the NS branch in the TF-QNNNNN case, both components of GW170817 are favored as NSs, and no NS is required to exceed $1.97\,M_\odot$. This relaxation of the maximum-mass constraint allows the NS matter EOS in the TF-QNNNNN case to be considerably softer than that in the NS case.
A comparison of Fig.~\ref{Fig-def-MR}(a) and~(b) indicates that the tidal information from GW170817 without the constraint $M>1.97\,M_\odot$ correspondingly favors smaller NS radii near $1.4\,M_\odot$, consistent with such a soft EOS.
Our Bayesian analysis yields $R_\text{1.4,TF-NS} = 11.16_{-0.42}^{+0.43}$~km and $M_\text{TOV,TF-NS} = 1.72_{-0.16}^{+0.18}\,M_\odot$ (68\% CI), both smaller than the corresponding values in the  NS case.
For the SQS branch, PSR J0740+6620 is the only CS among the six CSs that constrains the TF-SQS mass-radius relation.
As a consequence, the inferred values, $R_\text{1.4,TF-SQS} = 11.14_{-0.69}^{+0.97}$~km and $M_\text{TOV,TF-SQS} = 2.63_{-0.31}^{+0.36} M_\odot$ (68\% CI), carry larger uncertainties than those in the  udQS case.

Current multimessenger observations favor a soft EOS for NS matter in the TF-QNNNNN case, characterized by small NS radii around $ 1.4 M_\odot $ and a low NS maximum mass.
This soft EOS has a natural physical origin: it can be realized through the inclusion of $\Delta$ resonances alongside hyperons~\cite{Drago:2013fsa,Drago:2014oja,Drago:2015cea,Bombaci:2020vgw} or by using a symmetry energy softening around intermediate densities (see, e.g., Ref.~\cite{Li:2022okx,Ye:2024meg}).
While this softening of the NS matter EOS is essential to reproduce the small radii observed in the NS branch, it simultaneously imposes a stringent theoretical upper limit on the NS maximum mass.
Some theoretical estimates indicate that the maximum mass of the NS branch should have a relatively low value, typically $\sim 1.5-1.6M_\odot$~\cite{Drago:2013fsa}.
These estimates align well with our finding of $M_{\text{TOV, TF-NS}} < 1.9M_\odot$ ($68\%$ CI) and constitute an interesting feature of the TF-QNNNNN case: the limitation, that the NS branch cannot support massive objects, requires the existence of a second, stiffer branch of QSs to account for massive pulsars like PSR J0740+6620, thereby naturally resolving the tension between the observations of small radii (requiring soft EOS) and large maximum masses (requiring stiff EOS) for various CSs in the single-family scenarios, since the large-mass CSs are SQSs instead of NSs in the TF-QNNNNN case.

The coexistence of NSs and SQSs, especially that only the heaviest PSR J0740+6620 is favored as a SQS in the most probable case TF-QNNNNN, is physically plausible within the formation picture according to the TF scenario. Even if SQM is absolutely stable, the conversion of an NS into an SQS is not automatic because it requires a sufficiently large central strangeness fraction. In more massive CSs, the higher central density favors the appearance of hyperons and increases the associated strangeness fraction, thereby facilitating the formation of an SQM seed and the subsequent conversion into an SQS~\cite{Bombaci:2004mt,Bombaci:2016xuj,Drago:2015cea,Wiktorowicz:2017swq}. Within this formation picture, PSR J0740+6620---the only source considered here with a mass near $2\,M_\odot$---can more readily cross the conversion threshold and become an SQS, while the other five sources remain NSs.

Indeed, neither single-family NS or udQS scenario alone can accommodate the full set of observational constraints.
As shown in Fig.~\ref{Fig-def-MR}(a) and (b), although the udQS case adequately describes the massive PSR J0740+6620, it cannot account for the large radius of PSR J0030+0451.
The NS case, by contrast, matches the properties of PSR~J0030+0451 but fails to simultaneously reconcile the small radius of PSR J0614$-$3329 with the high mass of PSR J0740+6620.
These complementary failures, i.e., one case excelling where the other fails, point to a fundamental limitation of any single-family assumption, naturally motivating a two-branch structure.
Furthermore, as discussed in Ref.~\cite{Mauviard:2025dmd}, the tension between the soft EOS required by the small radius of PSR J0614$-$3329 and the stiff EOS required by the massive PSR J0740+6620 results in a bimodal structure in the posterior of $R_\text{1.4,NS}$ (see Fig. 9 in Ref.~\cite{Mauviard:2025dmd}).
This bimodal structure implies that the mass-radius relation may possess two distinct branches, a feature intrinsic to the TF scenario.
Consistent with this interpretation, the mass-radius posterior in the TF-QNNNNN case, which naturally contains two branches, passes through the central regions of all the multimessenger constraints.
Collectively, these results clearly demonstrate why the most favored TF case provides a superior global fit to the full suite of observational constraints compared to the NS and udQS cases.

Beyond the mass-radius relation, the tidal deformability offers a complementary probe of the CS interior.
As shown in Fig.~\ref{Fig-def-MR}(c) and (d), NSs and QSs exhibit markedly different $\Lambda$ values in the three cases, with the TF-NS branch yielding smaller deformabilities that better describe the GW170817 data.
Note that our Bayesian analysis uses the tidal information extracted directly from the GW waveform, rather than the tidal posterior distributions obtained after imposing additional EOS information~\cite{LIGOScientific:2018cki}.
In the TF-QNNNNN case, the tidal deformability of a $1.4M_\odot$ CS is inferred to be $\Lambda_{\text{1.4,TF-NS}} = 194_{-53}^{+68}$ for NSs and $\Lambda_{\text{1.4,TF-SQS}} = 567_{-195}^{+393}$ for QSs (68\% CI).
By comparison, we note the single-family NS and udQS cases respectively yield $\Lambda_{\text{1.4,NS}} = 296_{-65}^{+88}$ and $\Lambda_{\text{1.4,udQS}} = 509_{-87}^{+112}$, indicating that the difference between the NS and udQS cases is smaller than that between TF-NS and TF-SQS in the TF-QNNNNN case.

Consistent with the radius changing trend shown in Fig.~\ref{Fig-def-MR}(a) and (b), where NSs in the TF-QNNNNN case have smaller radii than those in the NS case,
one sees from Fig.~\ref{Fig-def-MR}(c) and (d) that
the GW170817 tidal data favor smaller NS tidal deformabilities in the TF-QNNNNN case.
Consequently, although the NS case remains compatible with the tidal constraints under the requirement $M_{\rm TOV} \geq 1.97M_\odot$~\cite{LIGOScientific:2018cki}, the TF-NS branch provides a better description of the original tidal data.
Notably, the udQSs (TF-SQSs) exhibit larger $ \Lambda $ values despite having smaller radii than the NSs (TF-NSs), a behavior attributed primarily to the sharp density discontinuity at the surface of QSs~\cite{Damour:2009vw,Hinderer:2009ca,Postnikov:2010yn,Takatsy:2020bnx}.
We note for both NSs and QSs in all cases, the established quasi-universal relations between $\Lambda$ and the dimensionless moment of inertia $\bar{I}$ (see, e.g., Ref.~\cite{Yagi:2016bkt}) remain well satisfied.
The Bayes factors of the three CS cases relative to the udQS case together with some key macroscopic quantities, including $M_{\rm TOV}$, $R_{1.4}$, $\Lambda_{1.4}$ for a $1.4M_\odot$ CS, and the central number density $n_c$ of the maximum-mass configuration are summarized in Tab.~\ref{Tab-summary-def}.

\begin{table}[htb!]
    \caption{Bayes factors of the NS and udQS cases, as well as TF-QNNNNN (the most favored case),  against the udQS case in the $\vec{d}_\text{def}$~(Note: $ \mathcal{Z}_\text{udQS} = 0.032 $). Properties of CSs in different cases are also included ($68\%$ CI).}
    \centering
    \setlength{\tabcolsep}{0.8pt}
	\footnotesize
    \begin{tabular}{|c|c|c|c|c|c|c|}
		\hline
        case& $\mathcal{B}^\text{case}_\text{udQS}$ & $M_{\rm TOV}$ ($M_\odot$) &  $R_\text{1.4}$ (km) & $\Lambda_\text{1.4}$ & $n_c / n_0$\\
		\hline
        NS & 1.0 & $2.09_{-0.08}^{+0.10}$ & $11.73_{-0.38}^{+0.40}$ & $296_{-65}^{+88}$ & $6.46_{-0.60}^{+0.58}$ \\
        \hline
        udQS &1 & $2.50_{-0.27}^{+0.30}$ & $10.99_{-0.29}^{+0.33}$ & $509_{-87}^{+112}$ & $6.38_{-0.96}^{+1.03}$ \\
        \hline
        \makecell{TF-QNNNNN \\(TF-NS)} &  \multirow{2}{*}{\makecell{6.1}} &  $1.72_{-0.16}^{+0.18}$ & $11.16_{-0.42}^{+0.43}$ & $194_{-53}^{+68}$&  $8.12_{-1.14}^{+1.38}$ \\
        \cline{1-1} \cline{3-6}
        \makecell{TF-QNNNNN \\(TF-SQS)} & & $2.63_{-0.31}^{+0.36}$ &$11.14_{-0.69}^{+0.97}$ & $567_{-195}^{+393}$ & $6.25_{-1.16}^{+1.16}$ \\
		\hline
    \end{tabular}
	\label{Tab-summary-def}
\end{table}

\begin{figure*}[!htbp]
	\centering
	\includegraphics[width=0.9\linewidth]{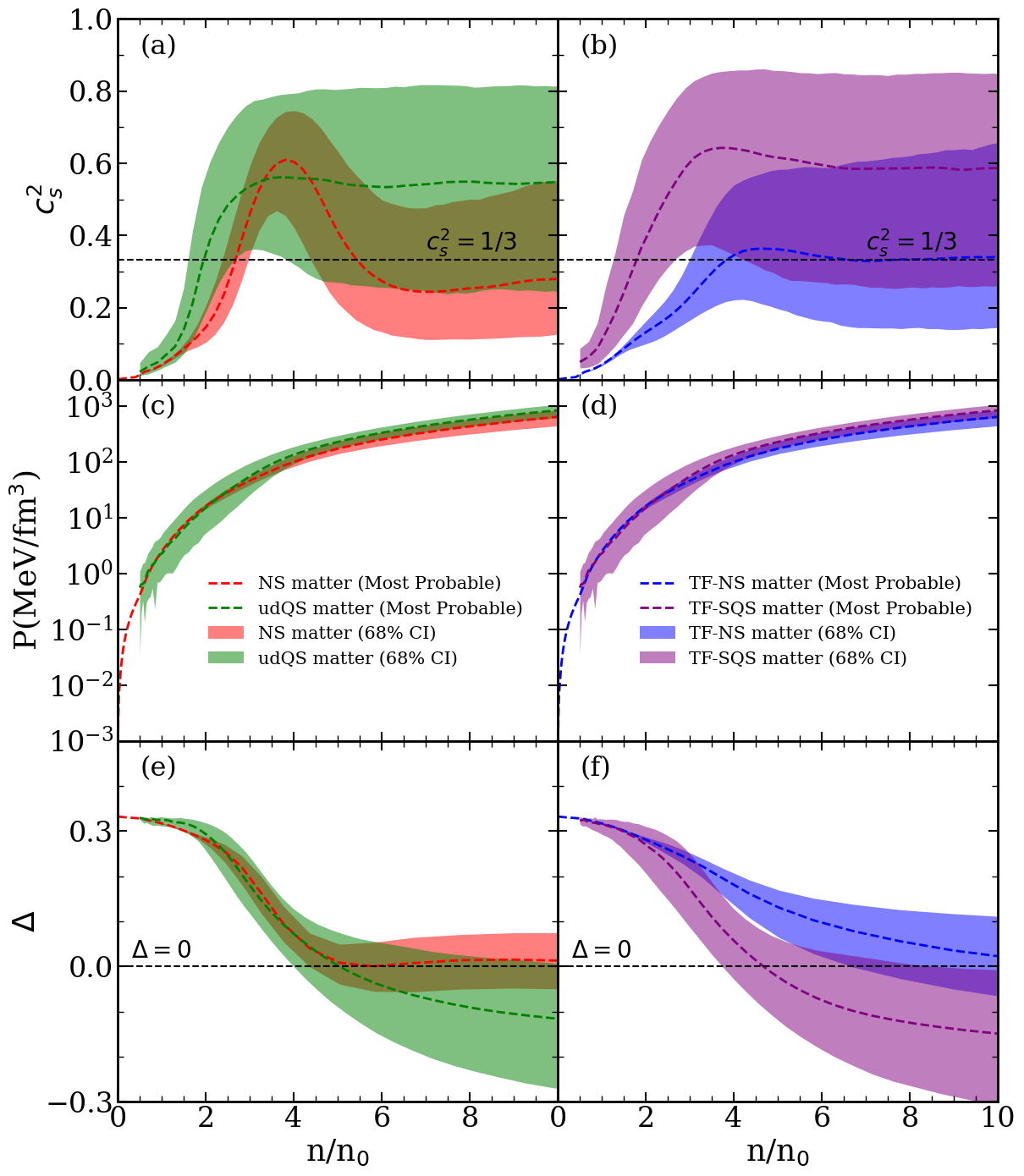}
	\caption{
        Posterior distributions (68\% CI) and the corresponding most probable value (dashed line) for the squared sound velocity $c_s^2$ (a), pressure $p$ (c), and trace anomaly $\Delta$ (e) in NS and QS matter are shown as functions of baryon number density in the NS and udQS cases.
        The corresponding results for the TF-QNNNNN are presented in panels (b), (d), and (f).
    }
	\label{Fig-def-EOS}
\end{figure*}

\subsubsection{Properties of dense matter}
We proceed to analyze the properties of dense matter in the three CS cases.
The sound velocity $c_s$ is an important quantity to characterize the EOS of dense matter.
Fig.~\ref{Fig-def-EOS}(a) illustrates the posterior distributions of $c_s^2$ of CS matter as a function of baryon number density for the NS and udQS cases.
One sees from Fig.~\ref{Fig-def-EOS}(a) that in the NS case, the squared sound speed $c_{\text{s,NS}}^2$ exhibits a pronounced peak reaching $c_{\text{s,NS,max}}^2 \sim 0.6 c^2$ at densities of $n \approx 4 n_0$. This non-monotonic feature has stimulated significant discussions regarding the effective degrees of freedom in dense matter~\cite{McLerran:2018hbz,Lee:2021hrw,Marczenko:2022jhl,Fujimoto:2022ohj,Pang:2023dqj}.
Such a peak could be interpreted as a signature of quarkyonic matter~\cite{McLerran:2018hbz}. Alternatively, it may arise from the high-density behavior of the symmetry energy~\cite{Zhang:2022sep,Wang:2023zcj} or signal the onset of hyperonization~\cite{Ye:2024meg}.
In contrast, as depicted in Fig.~\ref{Fig-def-EOS}(b), the squared speed of sound for NS matter in the TF-QNNNNN case, denoted $c_{\text{s,TF-NS}}^2$, rises monotonically and subsequently saturates at approximately $c^2/3$ for densities exceeding $n \approx 4n_0$.
This small value of the sound speed implies a softer EOS for NS matter within the TF-QNNNNN case, leading to increased compactness (see Fig.~\ref{Fig-def-MR}(b)) and a substantial central density of $n_{\text{c,TF-NS}} = 8.12_{-1.14}^{+1.38} n_0$ (68\% CI) for the maximum-mass case (see Tab.~\ref{Tab-summary-def}).
Within the cases considered here, the pronounced peak in $c_s^2$ appears only in the single-family NS case: as shown in Fig.~\ref{Fig-def-EOS}(a) and (b), neither $c_{\text{s,udQS}}^2$ nor $c_{\text{s,TF-SQS}}^2$ exhibits such a feature, both instead rising monotonically like $c_{\text{s,TF-NS}}^2$.
Taken together, the contrasting sound-speed profiles described above underscore that the evolution of $c_s^2$ at intermediate densities within CSs is highly sensitive to the choice of  CS scenario (as illustrated by comparing Fig.~\ref{Fig-def-EOS}(a) to Fig.~\ref{Fig-def-EOS}(b)) as well as the constraints from lower densities and massive CSs (as shown by the significantly different sound speed behaviors between NSs and QSs).

The pressure as a function of baryon density for NS and QS matter within the NS, udQS, and TF-QNNNNN cases is displayed in Fig.~\ref{Fig-def-EOS}(c) and (d).
The EOS of NS matter is stringently constrained by the BPS EOS and ChEFT calculations in the density region below $1.5n_0$.
Conversely, the pressure of QS matter at lower densities exhibits significantly larger uncertainties due to weak constraints at low densities.
In particular, the energy per baryon $\mu_1$ and the baryon number density $n_1$ for QS matter at the point of vanishing pressure are inferred (in 68\% CI) to be $\mu_1 = 693_{-105}^{+119}$~MeV and $n_1 = 2.15_{-0.62}^{+0.62} n_0$ within the TF-QNNNNN case, respectively, and $\mu_1 = 724_{-104}^{+111}$~MeV and $n_1 = 2.10_{-0.43}^{+0.50} n_0$ within the udQS case, respectively. It is interesting to see that these results are consistent with those in conventional quark star studies~\cite{Chu:2017wip,Cao:2020zxi,Miao:2021nuq}.

Figure~\ref{Fig-def-EOS}(e) and (f) show the trace anomaly normalized by the energy density, $ \Delta = 1/3 - p/\varepsilon $, which has recently been proposed as a new measure of conformality~\cite{Fujimoto:2022ohj}.
From Fig.~\ref{Fig-def-EOS}(e), it is evident that conformal symmetry may be restored above $\sim 5 n_0 $ in the NS case, which is consistent with recent findings by other groups~\cite{Fujimoto:2022ohj,Marczenko:2022jhl,Annala:2023cwx}.
In contrast, no such restoration is observed in either the TF-QNNNNN or udQS case.
As noted in Ref.~\cite{Fujimoto:2022ohj}, the squared sound speed can be decomposed into derivative and nonderivative contributions in terms of $\Delta$,
$c_{\mathrm{s}}^2 / c^2 = 1/3 - \Delta - \varepsilon d\Delta / d\varepsilon$,
and the peak in the sound speed observed in NS matter in the NS case can be attributed to the derivative term involving $\Delta$.
Furthermore, our results suggest that while NS matter in both the NS and TF-NS appears to satisfy the conjecture~\cite{Fujimoto:2022ohj} that the matter contribution to the trace anomaly is positive definite, QS matter in the udQS and TF-SQS  violates this conjecture.

\subsubsection{Implications for nuclear physics, astrophysics and cosmology}
Beyond the properties of dense matter, the preference of the TF-QNNNNN case carries profound implications for both nuclear physics and astrophysics, rooted in the coexistence of a soft TF-NS branch and a stiff TF-SQS branch.
For nuclear physics, the inferred $M_\text{TOV,TF-NS}=1.72_{-0.16}^{+0.18}\,M_\odot$ on the TF-NS branch provides a natural, data-driven resolution of the long-standing hyperon puzzle and the $\Delta$-resonance puzzle~\cite{Drago:2013fsa,Drago:2014oja,Drago:2015cea}.
In single-family frameworks, the onset of hyperons near $2$--$3\,n_0$ and that of $\Delta(1232)$ near $2\,n_0$ soften the hadronic EOS and preclude the NS branch from reaching $2\,M_\odot$, in tension with the heaviest known pulsars.
This tension is absent in the TF-QNNNNN case: hyperons and $\Delta$ resonances may freely populate the NS core, while the astrophysical observations with $M \gtrsim 2\,M_\odot$ can be explained entirely by the SQS branch.

On the astrophysical side, the large $M_\text{TOV,TF-SQS}=2.63_{-0.31}^{+0.36} M_\odot$ sheds light on several observational puzzles.
For the GW170817 merger remnant from a binary with total mass $M_\text{tot}\sim 2.73\,M_\odot$, the object could have survived as a stable or slowly rotating supramassive TF-SQS rather than promptly collapsing to a black hole, consistent with X-ray evidence for a long-lived remnant~\cite{Piro:2018bpl}.
Similarly, the secondary of GW190814 with mass $ M \sim 2.6\,M_\odot$~\cite{LIGOScientific:2020zkf}, the secondary of GW200210\_092254 with mass $ M \sim 2.83\,M_\odot$~\cite{KAGRA:2021vkt}, and the millisecond pulsar PSR~J0952$-$0607 with mass $ M \sim 2.35\,M_\odot$~\cite{Romani:2022jhd,Romani:2025ytn} can be all naturally explained as TF-SQSs, without invoking an extremely stiff hadronic EOS or near-Keplerian rotation.

More broadly, the absolute stability of SQM carries implications that extend into cosmology.
As first argued by Witten~\mbox{\cite{Witten:1984rs}}, the absolute
stability of SQM, which is required by the existence of self-bound SQSs,
implies that macroscopic SQM clusters (strangelets) could have formed in the early universe and survived to the present day (see Ref.~\cite{DiClemente:2024lzi} for the recent revisit of this topic).
Since the geometric cross section $\sigma $ of a strangelet scales as $A^{2/3}$
while its mass grows as $A$, the reduced cross-section $\sigma/M$, which controls the strength of dark-matter interactions on cosmological observables, scales as $A^{-1/3}$ and becomes vanishingly small at macroscopic baryon numbers.
Such strangelets would therefore be effectively weakly interacting
despite being baryonic, making them possible dark-matter candidates~\cite{DiClemente:2024lzi}.
The moderate evidence for TF-QNNNNN therefore motivates an intriguing, although indirect and speculative, connection between compact-star observations and strangelet dark matter.

\begin{figure*}[!htbp]
	\centering
	\includegraphics[width=0.95\linewidth]{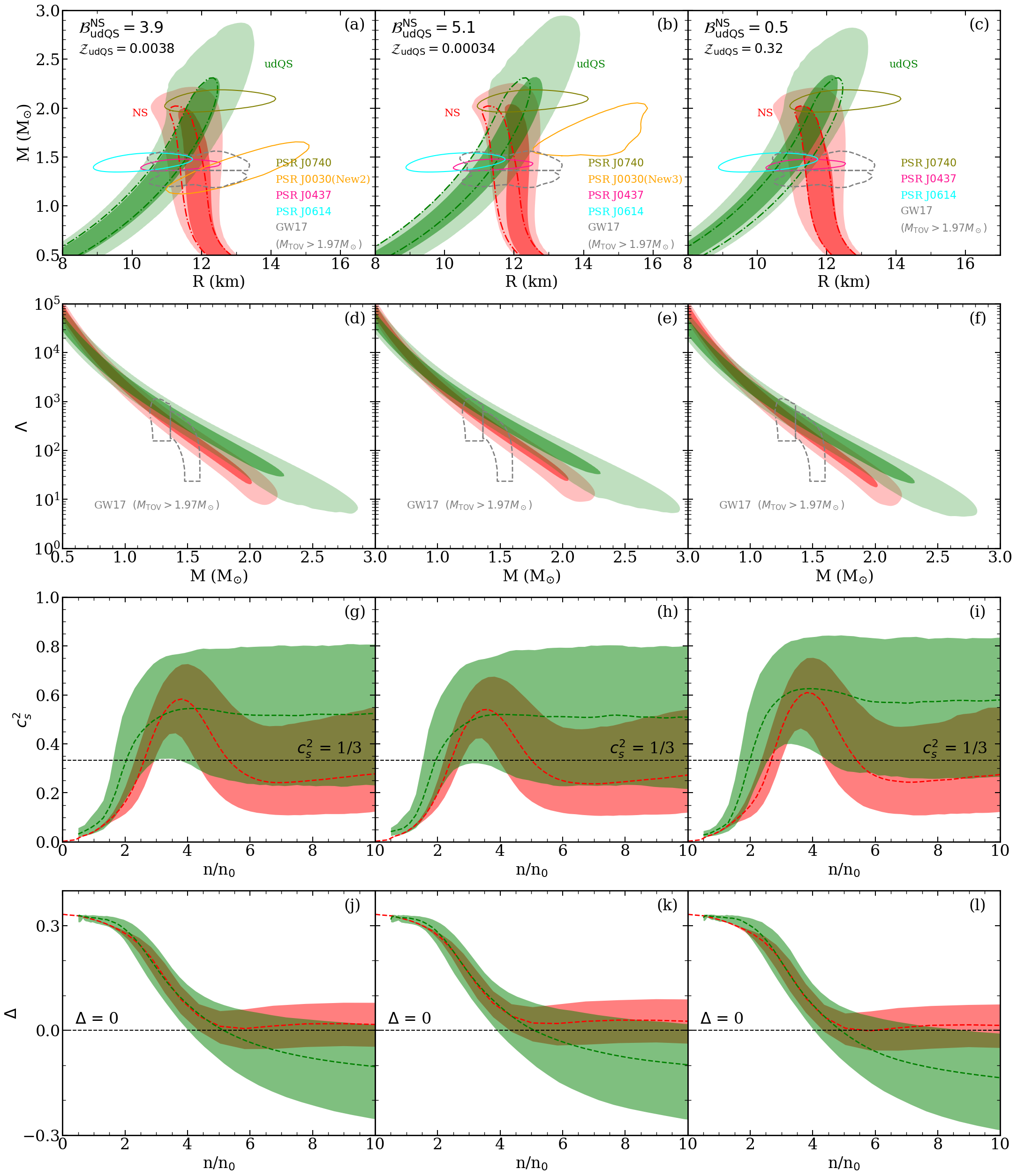}
	\caption{
    Properties of CSs and dense matter in different datasets: J0030New2 (left column), J0030New3 (middle column) and w/o-J0030 (right column).
    For comparison, we also show the mass-radius posterior distribution in 68\% CI obtained in the $\vec{d}_\text{def}$ (dash-dotted line).
    The other constraints included are identical to those adopted in Fig.~\ref{Fig-def-MR}.
    }
	\label{Fig-disJ0030-NSudQS}
\end{figure*}

\subsection{Impact of PSR J0030+0451}
\label{ImpactJ0030}
The NICER mass-radius determination for PSR J0740+6620, PSR J0437$-$4715, and PSR J0614$-$3329 is insensitive to the adopted hot spot modeling.
However, as discussed in Ref.~\cite{Rutherford:2024srk}, the NICER mass-radius relation of PSR J0030+0451 varies under different hot spot modeling choices.
In our default dataset $ \vec{d}_\text{def} $, the mass-radius relation of PSR J0030+0451 is obtained using the \texttt{ST+PDT} model~\cite{Vinciguerra:2023qxq}.
To assess how the choice of hot spot model affects our results, we consider the following three datasets:
(1) J0030New2, which is identical to $ \vec{d}_\text{def} $ but with PSR J0030+0451 analyzed using the \texttt{ST+PST} model~\cite{Vinciguerra:2023qxq};
(2) J0030New3, which is identical to $ \vec{d}_\text{def} $ but with PSR J0030+0451 analyzed using the \texttt{PDT-U} model~\cite{Vinciguerra:2023qxq};
(3) w/o-J0030, which removes PSR J0030+0451 from $ \vec{d}_\text{def} $.
As we show below, the most favored case belonging to TF scenario remains robust among all four datasets, and the \texttt{ST+PDT} model used in $ \vec{d}_\text{def} $ is most preferred by the multimessenger data and {\it ab initio} calculations.

\subsubsection{NS and udQS scenarios}
We first examine the impact of different hot spot modeling for PSR J0030+0451 in the NS and udQS cases.
Figure~\ref{Fig-disJ0030-NSudQS}(a) presents the posterior mass-radius distributions for NSs and udQSs obtained using the J0030New2 dataset.
In this dataset, the maximum mass of CS is estimated to be $ M_\text{TOV,NS} = 2.10_{-0.09}^{+0.11}M_\odot$ and $ M_\text{TOV,udQS} = 2.47_{-0.26}^{+0.32}M_\odot$, both consistent with the corresponding results in $\vec{d}_\text{def}$.
Furthermore, the inferred radii exhibit a slight increase relative to the results in  $\vec{d}_\text{def}$: $R_\text{1.4,NS}$ increases from $11.73_{-0.38}^{+0.40}$~km in $\vec{d}_\text{def}$ to $11.86_{-0.42}^{+0.37}$~km, and $R_\text{1.4,udQS}$ increases from $10.99_{-0.29}^{+0.33}$~km in $\vec{d}_\text{def}$ to $11.07_{-0.31}^{+0.38}$~km (68\% CI).
Although these shifts are relatively small, they lead to a substantial change in model preference: the Bayes factor favoring the NS case over the udQS case, $\mathcal{B}^{\text{NS}}_{\text{udQS}}$, rises notably from 1.0 for $\vec{d}_\text{def}$ to 3.9.
This behavior can be understood from Fig.~\ref{Fig-disJ0030-NSudQS}(a), namely, the inferred udQS mass-radius relation does not fit the constraints of PSR J0030+0451 in the J0030New2 dataset (note that the Bayesian evidence of the udQS case significantly decreases from $\mathcal{Z}_\text{udQS} = 0.032$ in $\vec{d}_\text{def}$ to $\mathcal{Z}_\text{udQS} = 0.0038$ in J0030New2), whereas the NS case remains consistent with the data.

With an even larger radius for PSR J0030+0451, the J0030New3 dataset further disfavors the udQS case.
As shown in Fig.~\ref{Fig-disJ0030-NSudQS}(b), the radii of both NSs and udQSs increase more substantially than in J0030New2 due to the even larger radius of PSR J0030+0451 in the J0030New3 dataset.
However, for J0030New3, the radius of PSR J0030+0451 becomes so large that the NS mass-radius relation only marginally overlaps with the left edge of the PSR J0030+0451 constraint at the 68\% CI, while the udQS mass-radius relation shows no overlap at all.
Consequently, the Bayes factor $\mathcal{B}^{\text{NS}}_{\text{udQS}}$ increases further from 1.0 for $ \vec{d}_\text{def} $  to 5.1 for J0030New3, while the Bayesian evidence of the udQS case becomes very small, i.e., $\mathcal{Z}_\text{udQS} = 0.00034$ in J0030New3.
Moreover, we obtain $R_\text{1.4,NS} = 12.11_{-0.39}^{+0.26}$~km and $ M_\text{TOV,NS} =  2.13_{-0.09}^{+0.11}M_\odot$, along with $R_\text{1.4,udQS} = 11.32_{-0.32}^{+0.37}$~km and $ M_\text{TOV,udQS} =  2.52_{-0.28}^{+0.35}M_\odot$ at the 68\% CI for J0030New3.

Taken together, the results from $\vec{d}_\text{def}$, J0030New2, and J0030New3 indicate that different mass-radius constraints for PSR J0030+0451 significantly influence model comparison.
To mitigate this source of systematic uncertainty, we remove the PSR~J0030+0451 constraint from the default dataset $\vec{d}_\text{def}$ (corresponding to the w/o-J0030 dataset) and repeat the Bayesian inference.
As shown in Fig.~\ref{Fig-disJ0030-NSudQS}(c), the inferred NS radius for w/o-J0030 remains essentially unchanged from that for $\vec{d}_\text{def}$.
This robustness may be due to the stringent requirement of a sufficiently stiff nuclear matter EOS imposed by the massive compact star PSR~J0740+6620, which in turn favors comparatively large radii around $ 1.4\,M_\odot$.
By contrast, the inferred udQS radius decreases modestly once the PSR~J0030+0451 constraint is removed, reflecting the fact that the udQS can support the high-mass constraint while still allowing relatively smaller radii.
For the w/o-J0030 dataset, we obtain $R_\text{1.4,NS} = 11.74_{-0.41}^{+0.43}$~km, $M_\text{TOV,NS} = 2.10_{-0.09}^{+0.11}M_\odot$, $R_\text{1.4,udQS} = 10.78_{-0.38}^{+0.34}$~km, and $M_\text{TOV,udQS} = 2.51_{-0.26}^{+0.26}M_\odot$ at the 68\% CI, with $\mathcal{B}^\text{NS}_\text{udQS}$ dropping from 1.0 in $\vec{d}_\text{def}$ to 0.5 in w/o-J0030 (Note the Bayesian evidence $\mathcal{Z}_\text{udQS}$ changes from $0.032$ in $\vec{d}_\text{def}$ to $0.32$ in w/o-J0030).

Figures~\ref{Fig-disJ0030-NSudQS}(d-f) show the tidal deformabilities of NSs and udQSs for the three datasets, i.e., J0030New2, J0030New3 and w/o-J0030, respectively.
For the NS and udQS cases, the tidal deformability increases in J0030New2 and J0030New3 compared to $\vec{d}_\text{def}$.
Specifically, we obtain
$ \Lambda_{\text{1.4,NS}} = 322_{-79}^{+88} $ and
$ \Lambda_{\text{1.4,udQS}} = 526_{-95}^{+126} $ in J0030New2,
and
$ \Lambda_{\text{1.4,NS}} = 379_{-86}^{+67} $ and
$ \Lambda_{\text{1.4,udQS}} = 606_{-103}^{+138} $ in J0030New3.
These results demonstrate that the tidal deformability becomes larger when a larger radius is adopted for PSR J0030+0451.
After removing PSR J0030+0451 from $ \vec{d}_\text{def} $, we find
$ \Lambda_{\text{1.4,NS}} = 298_{-73}^{+95} $ and
$ \Lambda_{\text{1.4,udQS}} = 449_{-95}^{+105} $ in w/o-J0030.

As for the properties of dense matter,
as illustrated in Fig.~\ref{Fig-disJ0030-NSudQS}(g-l), the choice of PSR J0030+0451 hot spot model does not significantly affect the results in either the NS or udQS case.
For the NS case, the squared speed of sound $c^2_{\text{s,NS}}$ retains a peak structure and approaches the conformal limit $c^2/3$ at high densities, with the corresponding trace anomaly $\Delta$ tending toward zero.
In contrast, the $c^2_{\text{s,udQS}}$ increases monotonically with density, reaching approximately $c^2/2$ above $\sim 3n_0$, while the corresponding trace anomaly $\Delta$ decreases steadily.

\begin{figure*}[!htbp]
	\centering
	\includegraphics[width=0.95\linewidth]{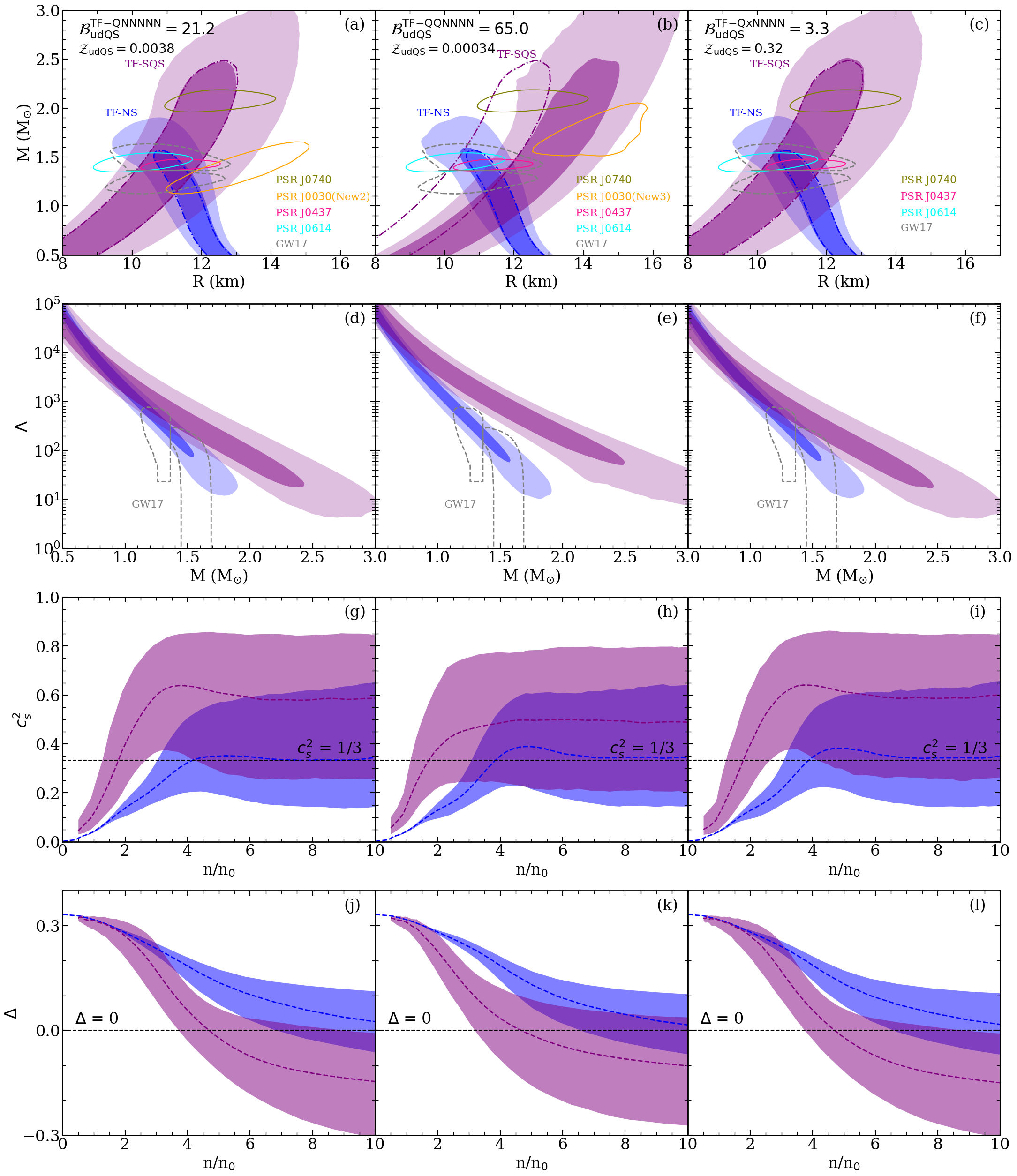}
	\caption{
    Similar to Figure~\ref{Fig-disJ0030-NSudQS}, but for the TF-QNNNNN(left), TF-QQNNNN(middle) and TF-QxNNNN(right).
    }
	\label{Fig-disJ0030-TF}
\end{figure*}

\subsubsection{TF scenario}
We now turn to the TF scenario, firstly comparing the J0030New2 results with those from the default $\vec{d}_\text{def}$.
Same as $\vec{d}_\text{def}$, we find the most favored case for J0030New2 remains TF-QNNNNN.
In this case, the most massive pulsar, PSR J0740+6620, is favored as a TF-SQS, while the other five CSs are favored as TF-NSs.
The mass-radius posteriors for the TF-NS and TF-SQS in this case are shown in Fig.~\ref{Fig-disJ0030-TF}(a), alongside the corresponding result from $\vec{d}_\text{def}$ for comparison.
One sees that the radius of the canonical TF-NS increases slightly compared to the result in $\vec{d}_\text{def}$, due to the larger radius of PSR J0030+0451 in the J0030New2 dataset, rising from $ R_\text{1.4,TF-NS} =11.16_{-0.42}^{+0.43}$ km in $\vec{d}_\text{def}$ to $ R_\text{1.4,TF-NS} = 11.20_{-0.46}^{+0.44}$ km in J0030New2.
Moreover, we note changing $\vec{d}_\text{def}$ to J0030New2 has a negligible impact on the maximum mass, yielding $ M_\text{TOV,TF-NS} = 1.71_{-0.15}^{+0.18}M_\odot $ in J0030New2, consistent with the value of $1.72_{-0.16}^{+0.18}M_\odot$ obtained from $\vec{d}_\text{def}$.
Since the constraints on the TF-SQS remain effectively unchanged, its mass-radius relation shows minor variation from the $\vec{d}_\text{def}$ result, with observed differences attributable to statistical fluctuations.
In particular, we obtain $R_\text{1.4,TF-SQS} = 11.15_{-0.69}^{+0.98}$~km and $ M_\text{TOV,TF-SQS} =  2.62_{-0.31}^{+0.36}M_\odot$ at the 68\% CI for J0030New2.

In contrast, we note that the results for the J0030New3 dataset change notably relative to $\vec{d}_\text{def}$.
In particular, the most favored case for J0030New3 changes to TF-QQNNNN,
namely, both the most massive objects PSR J0740+6620 and PSR J0030+0451 with the largest radius, are favored as TF-SQSs.
As illustrated in Fig.~\ref{Fig-disJ0030-TF}(b), to describe these two stars simultaneously, the radius of the TF-SQS increases substantially compared to the result in $\vec{d}_\text{def}$, intersecting the lower radius boundary of PSR J0030+0451 and the upper radius boundary of PSR J0740+6620.
Consequently, $ R_\text{1.4,TF-SQS} $ increases from $11.14_{-0.69}^{+0.97}$~km in $ \vec{d}_\text{def} $ to $12.74_{-0.72}^{+0.78}$~km in J0030New3, and $M_\text{TOV,TF-SQS} $ increases from $2.63_{-0.31}^{+0.36} M_\odot $ in $\vec{d}_\text{def}$ to $ 2.84_{-0.50}^{+0.56} M_\odot $ in J0030New3.
From Figs.~\ref{Fig-disJ0030-NSudQS} and~\ref{Fig-disJ0030-TF}, one can see that the case TF-QQNNNN in the TF scenario is perhaps the only choice that can reasonably describe the J0030New3 dataset.
In addition, we obtain $R_\text{1.4,TF-NS} = 11.05_{-0.45}^{+0.46}$~km and $ M_\text{TOV,TF-NS} = 1.72_{-0.15}^{+0.17}M_\odot$ at the 68\% CI for the J0030New3 dataset.

Using the w/o-J0030 dataset, which removes PSR J0030+0451 from $\vec{d}_\text{def}$, we repeat the previous analyses.
We find the most favored case becomes TF-QxNNNN, where `x' denotes the exclusion of PSR J0030+0451.
As shown in Fig.~\ref{Fig-disJ0030-TF}(c),
the removal of PSR J0030+0451 leaves the mass-radius relations for both the TF-NS and TF-SQS essentially unaffected
compared to the results in $\vec{d}_\text{def}$.
In particular, for w/o-J0030, we obtain $R_\text{1.4,TF-NS} = 11.05_{-0.46}^{+0.45}$~km and $ M_\text{TOV,TF-NS} =  1.72_{-0.15}^{+0.17} M_\odot$, along with $R_\text{1.4,TF-SQS} = 11.14_{-0.68}^{+0.96} $~km and $ M_\text{TOV,TF-SQS} =  2.63_{-0.31}^{+0.35} M_\odot$ at the 68\% CI.
Although the global properties of TF-QxNNNN in w/o-J0030 do not change significantly compared to the results in $ \vec{d}_\text{def}$, $\mathcal{B}^{\text{TF-QxNNNN}}_{\text{udQS}}$ decreases to $3.3$, compared to $6.1$ for $\vec{d}_\text{def}$.
This decrease in the Bayes factor occurs because the udQS case describes the w/o-J0030 dataset better than it does in $\vec{d}_\text{def}$ (Note $ \mathcal{Z}_\text{udQS} = 0.32$ for w/o-J0030, see, e.g., Fig.~\ref{Fig-disJ0030-NSudQS}(c), while $ \mathcal{Z}_\text{udQS} = 0.032$ for $\vec{d}_\text{def}$, see, e.g.,  Fig.~\ref{Fig-def-MR}(a)),
though the TF-QxNNNN describes the w/o-J0030 dataset as well as it does in $ \vec{d}_\text{def} $.
Consequently, the relative statistical advantage of the TF-QxNNNN  decreases.
This feature underscores that Bayes factors quantify the relative preference between competing hypotheses rather than providing an absolute measure for any single case.

In Figs.~\ref{Fig-disJ0030-TF}(d-f), we present the posterior distributions of the tidal deformability for the cases of TF-QNNNNN, TF-QQNNNN, and TF-QxNNNN in the J0030New2, J0030New3, and w/o-J0030 datasets, respectively.
The inferred tidal deformabilities for a $1.4M_\odot$ star reveal distinct difference for the TF-NS and TF-SQS.
For the TF-NS, the inferred values are $198_{-58}^{+74}$, $179_{-51}^{+69}$, and $179_{-53}^{+69}$ for J0030New2, J0030New3 and w/o-J0030, respectively.
These results are fully consistent with the constraints derived from the binary neutron star merger GW170817, which requires $\Lambda_{1.4} \lesssim 580$~\cite{LIGOScientific:2018cki}.
Conversely, the TF-SQS exhibits significantly larger tidal deformabilities: $567_{-197}^{+402}$, $1308_{-401}^{+586}$, and $564_{-192}^{+392}$ for J0030New2, J0030New3 and w/o-J0030, respectively.
Since tidal deformability scales strongly with stellar radius ($\Lambda \propto R^5$), the large radius at $1.4M_\odot$ in J0030New3 results in a rather large tidal deformability.

For the properties of dense matter in the TF scenario,
as illustrated in Figs.~\ref{Fig-disJ0030-TF}(g-l), the choice of the hot spot model for PSR J0030+0451 does not significantly affect the results.
For all the PSR J0030+0451 hot spot models~(i.e., the J0030New2, J0030New3, and w/o-J0030 datasets), one sees from Figs.~\ref{Fig-disJ0030-TF}(g-i) that the squared speed of sound for NS~(SQS) matter in the TF scenario rises monotonically and subsequently
saturates at approximately $c^2/3$~($0.6c^2$) for densities exceeding $n \approx 4n_0$, and no obvious peak is observed.
In addition, as shown in Figs.~\ref{Fig-disJ0030-TF}(j-l),  conformal symmetry restoration is not observed in the  TF scenario. These results are similar to those in $\vec{d}_\text{def}$ as shown in Figs.~\ref{Fig-def-EOS}(b) and (f).

\begin{table}[htb!]
    \centering
    \setlength{\tabcolsep}{0.4pt}
	\footnotesize
    \caption{Bayes factors of the NS and udQS cases, as well as of the most favored case within TF scenario, against the udQS case in different datasets.
    (Note: $ \mathcal{Z}_\text{udQS} = 0.0038$ for J0030New2,
    $ \mathcal{Z}_\text{udQS} = 0.00034$ for J0030New3, and
    $ \mathcal{Z}_\text{udQS} = 0.32$ for w/o-J0030.)
    Properties of CSs in different cases are also shown in 68\% CI.}
    \begin{tabular}{|c|c|c|c|c|c|c|}
		\hline

        & case&  $\mathcal{B}^\text{case}_\text{udQS}$ & { $M_{\rm TOV}$($M_\odot$)} &  $R_\text{1.4}$(km) & $\Lambda_\text{1.4}$ & $n_c / n_0$\\

        \hline
        \multirow{4}{*}{\makecell{J0030\\ New2}} &  NS &  3.9 & $2.10_{-0.09}^{+0.11}$ & $11.86_{-0.42}^{+0.37}$ & $322_{-79}^{+88}$ &  $6.34_{-0.62}^{+0.64}$ \\
        \cline{2-7}
        &  udQS & 1& $2.47_{-0.26}^{+0.32}$ & $11.07_{-0.31}^{+0.38}$ & $526_{-95}^{+126}$ & $6.38_{-0.98}^{+1.05}$ \\
        \cline{2-7}
        & \makecell{\tiny TF-QNNNNN \\(TF-NS)} & \multirow{2}{*}{\makecell{21.2}}  & $1.71_{-0.15}^{+0.18}$ &  $11.20_{-0.46}^{+0.44}$ & $198_{-58}^{+74}$ & $8.19_{-1.23}^{+1.43}$ \\
        \cline{2-2}\cline{4-7}
        & \makecell{\tiny TF-QNNNNN \\(TF-SQS)} & & $2.62_{-0.31}^{+0.36}$ & $11.15_{-0.69}^{+0.98}$ & $567_{-197}^{+402}$ & $6.24_{-1.17}^{+1.17}$ \\
        \hline

        \multirow{4}{*}{\makecell{J0030\\ New3}} &  NS & 5.1 & $2.13_{-0.09}^{+0.11}$ & $12.11_{-0.39}^{+0.26}$ & $379_{-86}^{+67}$ & $6.02_{-0.55}^{+0.64}$ \\
        \cline{2-7}
        &  udQS &1 & $2.52_{-0.28}^{+0.35}$ & $11.32_{-0.32}^{+0.37}$ & $606_{-103}^{+138}$ & $6.12_{-0.95}^{+1.04}$ \\
        \cline{2-7}

        & \makecell{\tiny TF-QQNNNN \\(TF-NS)} &  \multirow{2}{*}{\makecell{65.0}}  &$1.72_{-0.15}^{+0.17}$ & $11.05_{-0.45}^{+0.46}$ & $179_{-51}^{+69}$ & $8.34_{-1.15}^{+1.33}$ \\
        \cline{2-2}\cline{4-7}
        & \makecell{\tiny TF-QQNNNN \\(TF-SQS)} & & $2.84_{-0.50}^{+0.56}$ & $12.74_{-0.72}^{+0.78}$ & $1308_{-401}^{+586}$ & $5.07_{-1.05}^{+1.18}$ \\


        \hline
        \multirow{4}{*}{\makecell{w/o- \\ J0030}}  & NS  & 0.5 & $2.10_{-0.09}^{+0.11}$ & $11.74_{-0.41}^{+0.43}$ & $298_{-73}^{+95}$ & $6.44_{-0.64}^{+0.61}$ \\
        \cline{2-7}
        &  udQS & 1 & $2.51_{-0.26}^{+0.26}$ & $10.78_{-0.38}^{+0.34}$ & $449_{-95}^{+105}$ & $6.59_{-0.94}^{+0.99}$ \\
        \cline{2-7}

        & \makecell{\tiny TF-QxNNNN \\(TF-NS)} & \multirow{2}{*}{\makecell{3.3}} & $1.72_{-0.15}^{+0.17}$ &  $11.05_{-0.46}^{+0.45}$ & $179_{-53}^{+69}$ & $8.34_{-1.18}^{+1.39}$ \\
        \cline{2-2}\cline{4-7}
        & \makecell{\tiny TF-QxNNNN \\(TF-SQS)} & &$2.63_{-0.31}^{+0.35}$ & $11.14_{-0.68}^{+0.96}$ & $564_{-192}^{+392}$ & $6.23_{-1.12}^{+1.20}$ \\
        \hline

    \end{tabular}
    \label{Tab-summary-J0030}
\end{table}

Finally, an interesting question is which hot spot model for PSR J0030+0451 is most compatible with the other astrophysical and theoretical constraints in $\vec{d}_\text{def}$.
The three datasets considered here differ solely in the hot spot model adopted for PSR J0030+0451, namely $(\texttt{ST+PDT}, \texttt{ST+PST}, \texttt{PDT-U})$ in $(\vec{d}_\text{def}, \text{J0030New2}, \text{J0030New3})$.
The corresponding Bayesian evidences are $(0.032, 0.0038, 0.00034)$ for the udQS case, $(0.032, 0.015, 0.0017)$ for the NS case, and $(0.20, 0.081, 0.022)$ for the most favored case (TF-QNNNNN in $\vec{d}_\text{def}$, TF-QNNNNN in J0030New2, TF-QQNNNN in J0030New3).
This monotonic decrease of Bayesian evidence from $\vec{d}_\text{def}$ to J0030New2 then to J0030New3 identifies \texttt{ST+PDT} (in $\vec{d}_\text{def}$) as the hot spot model most strongly supported by the joint multimessenger constraints together with state-of-the-art \textit{ab initio} ChEFT and pQCD calculations.
This feature implies that the results obtained in the $\vec{d}_\text{def}$ dataset are relatively more consistent compared to the results in the datasets of J0030New2 and J0030New3.

The Bayes factors of various cases relative
to the corresponding udQS case for different hot spot models of PSR J0030+0451, together with some key macroscopic
quantities are summarized in Tab.~\ref{Tab-summary-J0030}.

Furthermore, we examine the dependence of the model-averaged Bayes factor of the TF scenario on the treatment of PSR J0030+0451.
For the datasets $(\vec{d}_\text{def}, \text{J0030New2}, \text{J0030New3}, \text{w/o-J0030})$, the model-averaged Bayes factors of the TF scenario (with an equal prior weight of $1/32$ for the 32 cases in w/o-J0030) relative to the corresponding udQS cases are $(1.2, 3.7, 5.3, 0.9)$. These values are lower than the Bayes factors $(6.1, 21.2, 65.0, 3.3)$ for the most favored individual cases because the average also includes the less favored assignments with equal prior weights. Combining the model-averaged TF Bayes factors with the corresponding Bayes factors of the NS case relative to the udQS case, $(1.0, 3.9, 5.1, 0.5)$, we find that the most favored scenarios within the Bayesian model-averaging framework are the TF, NS, TF, and udQS scenarios, respectively. This variation indicates that the scenario preference is sensitive to the adopted treatment of PSR J0030+0451.

Bayesian model averaging also allows us to assess the nature of each individual CS without relying on a single case.
With an equal prior weight for each case in the TF scenario, the probability for the $i$-th CS to be a QS is $P_i(\text{Q}) = \sum_{a:\,i\in\text{Q}} \mathcal{B}^a_\text{udQS} \big/ \sum_{a} \mathcal{B}^a_\text{udQS}$, where the numerator runs over the cases in which the $i$-th CS is a QS.
For the datasets $(\vec{d}_\text{def}, \text{J0030New2}, \text{J0030New3}, \text{w/o-J0030})$, the probabilities for the CSs to be QSs are $(82\%, 77\%, 67\%, 80\%)$ for PSR J0740+6620, $(35\%, 17\%, 43\%, -)$ for PSR J0030+0451, $(40\%, 37\%, 33\%, 39\%)$ for PSR J0437$-$4715, $(50\%, 53\%, 47\%, 52\%)$ for PSR J0614$-$3329, $(36\%, 38\%, 34\%, 38\%)$ for GW17\_1, and $(37\%, 40\%, 35\%, 40\%)$ for GW17\_2.
The probability for PSR J0740+6620 to be an SQS is above $60\%$ in all datasets, indicating that its identification as an SQS is robust against the treatment of PSR J0030+0451 and is considerably better established than the classification of the other five CSs.
For the other five CSs, the probabilities to be QSs are all below $55\%$. In particular, PSR J0030+0451, PSR J0437$-$4715, and the two components of GW170817 are more likely NSs than QSs in all datasets, with probabilities to be QSs below $45\%$, while PSR J0614$-$3329 has a probability close to $50\%$ and its nature cannot be determined by current data.

\section{Conclusion and Outlook}
In this work, we have performed a systematic, physics-agnostic Bayesian inference of the properties of dense matter and compact stars (CSs) by combining  multimessenger astrophysical data with the state-of-the-art \textit{ab initio} theoretical constraints.
Employing a nonparametric Gaussian process regression method to construct the equation of state~(EOS) with minimal model dependence, we have systematically compared 66 competing cases spanning three scenarios for the nature of CSs---the neutron star (NS) scenario, the up-down quark star (udQS) scenario, and the two-family (TF) scenario---and find that
TF-QNNNNN is the most favored case, with moderate evidence relative to the NS and udQS cases. In this case, PSR J0740+6620 is assigned to the strange quark star (SQS) branch and the other five CSs are assigned to the NS branch.

Our analysis, based on the mass-radius measurements from NICER for PSRs J0740+6620, J0030+0451, J0437$-$4715, and J0614$-$3329, tidal deformability constraints from GW170817, and  bounds from chiral effective field theory (ChEFT) and perturbative QCD (pQCD), yields the following key findings:

\begin{itemize}
\item \textbf{The Most Favored Case Belongs to the Two-Family Scenario:} Systematic Bayesian model selection among all 66 competing cases identifies TF-QNNNNN---belonging to the TF scenario---as the most favored, with a Bayes factor of $6.1$ relative to both the NS and udQS cases. This suggests that the astrophysical CS population may consist of two distinct branches, i.e., the NS branch and the QS branch. We would like to point out that the value $6.1$ refers to the single most favored assignment among the 64 cases in the TF scenario. When the evidence is averaged over all assignments, the Bayes factor of the TF scenario as a whole relative to the udQS case is $1.2$.

{\item \textbf{Nature of Individual Compact Stars:} Within the most probable case of the TF scenario, the massive pulsar PSR J0740+6620 ($M \sim 2\,M_\odot$) is favored as a strange quark star. In contrast, the lower-mass CSs---including the two components of GW170817 and the other NICER pulsars (PSR J0030+0451, J0437$-$4715, J0614$-$3329)--- are favored as neutron stars. This NS--SQS classification naturally accommodates very massive ($\sim 2.6\,M_\odot$) SQSs alongside a population of NSs whose maximum mass is correspondingly lower ($\sim 1.7\,M_\odot$).
}
\item \textbf{Properties of Dense Matter:} The inferred EOSs and their derived CS properties differ markedly between cases:
    \begin{itemize}
\item \textbf{Maximum Masses:} In the most favored case TF-QNNNNN, the maximum mass of non-rotating CSs is found to be $M_{\text{TOV}} = 1.72^{+0.18}_{-0.16} \, M_\odot$ for NSs and $M_{\text{TOV}} = 2.63^{+0.36}_{-0.31} \, M_\odot$ for SQSs (68\% CI).
    \item \textbf{Speed of Sound and Trace Anomaly:} Different density dependences are inferred for the sound speed ($c_s$). In the most favored case TF-QNNNNN, $c_s$ in both NS and SQS matter increases monotonically with density, eventually saturating at high densities.  This monotonic rise followed by saturation is different from the pronounced peak in $c_s$ often found in analyses assuming the NS case alone.  The trace anomaly also exhibits a smoother evolution in the TF case.
    \end{itemize}

\item \textbf{Robustness and Implications:} We have tested the robustness of these conclusions by varying the NICER hot-spot model for PSR J0030+0451. The Bayes factors and the favored classification of PSR J0030+0451 depend on the treatment of this source, while PSR J0740+6620 continues to be favored as an SQS in the most favored case.
\end{itemize}

More broadly, our results motivate possible implications for QCD theory, dark-matter physics, and compact-star classification:
\begin{itemize}
\item \textbf{Evidence for Absolute Stability of Strange Quark Matter:} The moderate Bayesian evidence for TF-QNNNNN, in which self-bound SQSs coexist with NSs, provides potential support for the Witten hypothesis that SQM may be the true ground state of strongly interacting matter.%

\item \textbf{Connection to Dark Matter:}
Under the Witten hypothesis, macroscopic strangelets could have formed in the early universe and survived to the present day. Their reduced interaction cross-section makes strangelet dark matter an intriguing, albeit speculative, possible connection between CS astrophysics and cosmology.

\item \textbf{New Framework for Compact Star Identification:}  The Bayesian model-selection framework developed here establishes a model-independent method for statistically assessing the nature of individual CSs based on multimessenger data. Precise mass-radius measurements from next-generation X-ray missions and detections of gravitational waves and electromagnetic signals from binary mergers involving CS components will be particularly valuable for sharpening the classification of individual CSs within this framework.

\end{itemize}

In conclusion, the multimessenger era is unveiling a more complex picture of the compact star population than previously assumed.
Our analysis favors a case in which NSs and SQSs coexist and PSR J0740+6620 is favored as an SQS, although the current evidence remains moderate.
Such a two-family picture offers a possible resolution of tensions among the astrophysical constraints and opens a new window into the phases of dense QCD matter.
Future observations from next-generation X-ray telescopes (e.g., eXTP~\cite{eXTP:2018anb,Li:2025uaw}) and gravitational-wave detectors (e.g. Einstein Telescope~\cite{Punturo:2010zz}) will be instrumental in confirming this paradigm and precisely constraining the dense matter EOS within both stellar families.

\section*{Acknowledgments}
The authors would like to thank Tyler Gorda, Sophia Han, Aleksi Kurkela, Ang Li, Yifeng Sun, Renxin Xu, Zhen Zhang, and Zhenyu Zhu for useful discussions.
This work was supported in part by the National Natural Science Foundation of China under Grant Nos. 12235010, the National SKA Program of China No. 2020SKA0120300, and the Science and Technology Commission of Shanghai Municipality under Grant No. 23JC1402700. ZC was supported by the T.D. Lee scholarship.
The computations in this paper were run on the Siyuan-1 cluster supported by the Center for High Performance Computing at Shanghai Jiao Tong University.

\bibliography{ref}

\end{document}